\documentclass[twoside,twocolumn,9pt]{article}
\usepackage{extsizes}
\usepackage[super,sort&compress,comma]{natbib}
\usepackage[version=3]{mhchem}
\usepackage[left=1.5cm, right=1.5cm, top=1.785cm, bottom=2.0cm]{geometry}
\usepackage{balance}
\usepackage{mathptmx}
\usepackage{sectsty}
\usepackage{graphicx}
\usepackage{lastpage}
\usepackage[format=plain,justification=justified,singlelinecheck=false,font={stretch=1.125,small,sf},labelfont=bf,labelsep=space]{caption}
\usepackage{float}
\usepackage{fancyhdr}
\usepackage{fnpos}
\usepackage[english]{babel}
\addto{\captionsenglish}{%
 
}
\usepackage{array}
\usepackage{droidsans}
\usepackage{charter}
\usepackage[T1]{fontenc}
\usepackage[usenames,dvipsnames]{xcolor}
\usepackage{setspace}
\usepackage[compact]{titlesec}
\usepackage{hyperref}

\usepackage{float}
\usepackage{scalerel}

\definecolor{cream}{RGB}{222,217,201}

\begin{document}

\pagestyle{fancy}
\thispagestyle{plain}
\fancypagestyle{plain}{
\renewcommand{\headrulewidth}{0pt}
}

\makeFNbottom
\makeatletter
\renewcommand\LARGE{\@setfontsize\LARGE{15pt}{17}}
\renewcommand\Large{\@setfontsize\Large{12pt}{14}}
\renewcommand\large{\@setfontsize\large{10pt}{12}}
\renewcommand\footnotesize{\@setfontsize\footnotesize{7pt}{10}}
\makeatother

\renewcommand{\thefootnote}{\fnsymbol{footnote}}
\renewcommand\footnoterule{\vspace*{1pt}%
\color{cream}\hrule width 3.5in height 0.4pt \color{black}\vspace*{5pt}}
\setcounter{secnumdepth}{5}

\makeatletter
\renewcommand\@biblabel[1]{#1}
\renewcommand\@makefntext[1]%
{\noindent\makebox[0pt][r]{\@thefnmark\,}#1}
\makeatother
\renewcommand{\figurename}{\small{Fig.}~}
\sectionfont{\sffamily\Large}
\subsectionfont{\normalsize}
\subsubsectionfont{\bf}
\setstretch{1.125} 
\setlength{\skip\footins}{0.8cm}
\setlength{\footnotesep}{0.25cm}
\setlength{\jot}{10pt}
\titlespacing*{\section}{0pt}{4pt}{4pt}
\titlespacing*{\subsection}{0pt}{15pt}{1pt}

\fancyfoot{}
\fancyfoot[LO,RE]{\vspace{-7.1pt}\includegraphics[height=9pt]{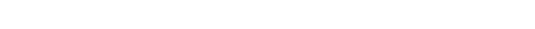}}
\fancyfoot[CO]{\vspace{-7.1pt}\hspace{13.2cm}\includegraphics{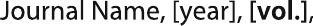}}
\fancyfoot[CE]{\vspace{-7.2pt}\hspace{-14.2cm}\includegraphics{head_foot/RF}}
\fancyfoot[RO]{\footnotesize{\sffamily{1--\pageref{LastPage} ~\textbar \hspace{2pt}\thepage}}}
\fancyfoot[LE]{\footnotesize{\sffamily{\thepage~\textbar\hspace{3.45cm} 1--\pageref{LastPage}}}}
\fancyhead{}
\renewcommand{\headrulewidth}{0pt}
\renewcommand{\footrulewidth}{0pt}
\setlength{\arrayrulewidth}{1pt}
\setlength{\columnsep}{6.5mm}
\setlength\bibsep{1pt}

\makeatletter
\newlength{\figrulesep}
\setlength{\figrulesep}{0.5\textfloatsep}

\newcommand{\topfigrule}{\vspace*{-1pt}%
\noindent{\color{cream}\rule[-\figrulesep]{\columnwidth}{1.5pt}} }

\newcommand{\botfigrule}{\vspace*{-2pt}%
\noindent{\color{cream}\rule[\figrulesep]{\columnwidth}{1.5pt}} }

\newcommand{\dblfigrule}{\vspace*{-1pt}%
\noindent{\color{cream}\rule[-\figrulesep]{\textwidth}{1.5pt}} }

\makeatother

\twocolumn[
 \begin{@twocolumnfalse}
{\includegraphics[height=30pt]{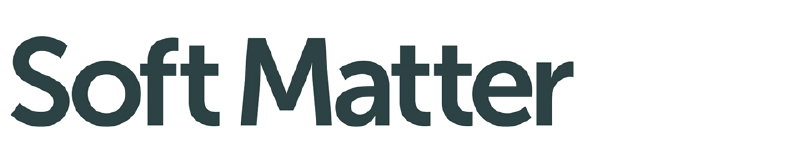}\hfill\raisebox{0pt}[0pt][0pt]{\includegraphics[height=55pt]{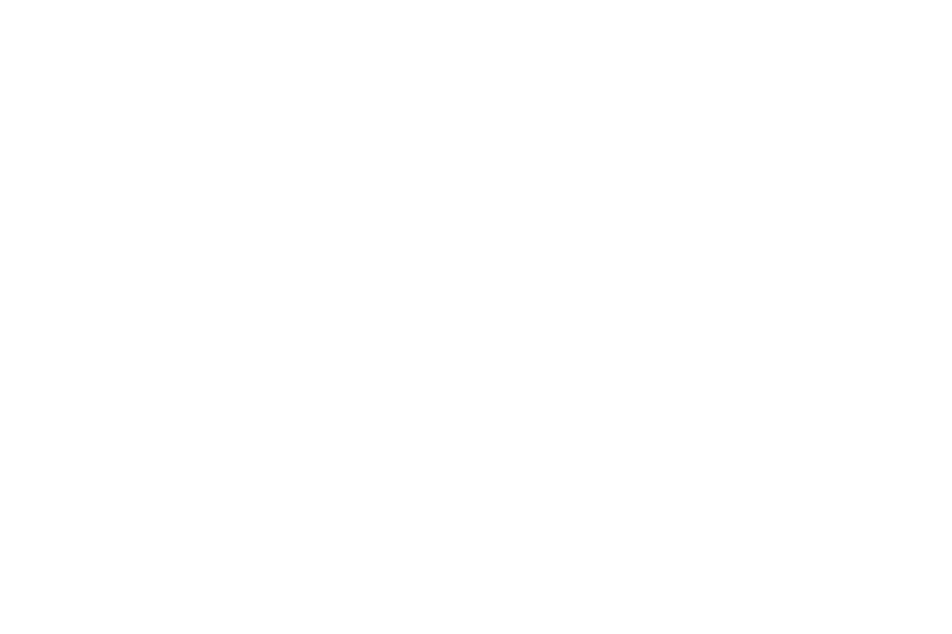}}\\[1ex]
\includegraphics[width=18.5cm]{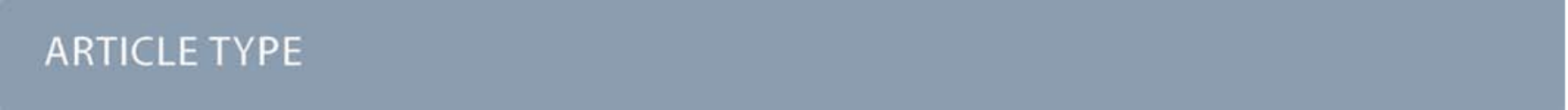}}\par
\vspace{1em}
\sffamily
\begin{tabular}{m{4.5cm} p{13.5cm} }

\includegraphics{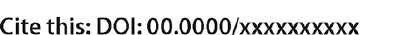} & \noindent\LARGE{\textbf{Coarse-grained theory for motion of solitons and skyrmions in liquid crystals}} \\
\vspace{0.3cm} & \vspace{0.3cm} \\

 & \noindent\large{Cheng Long\textit{$^{a}$} and Jonathan V. Selinger$^{\ast}$\textit{$^{a}$}} \\

\includegraphics{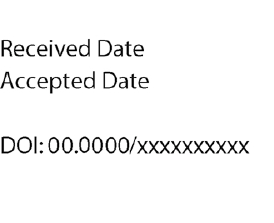} & \noindent\normalsize{Recent experiments have found that applied electric fields can induce motion of skyrmions in chiral nematic liquid crystals. To understand the magnitude and direction of the induced motion, we develop a coarse-grained approach to describe dynamics of skyrmions, similar to our group’s previous work on the dynamics of disclinations. In this approach, we represent a localized excitation in terms of a few macroscopic degrees of freedom, including the position of the excitation and the orientation of the background director. We then derive the Rayleigh dissipation function, and hence the equations of motion, in terms of these macroscopic variables. We demonstrate this theoretical approach for 1D motion of a sine-Gordon soliton, and then extend it to 2D motion of a skyrmion. Our results show that skyrmions move in a direction perpendicular to the induced tilt of the background director. When the applied field is removed, skyrmions move in the opposite direction but not with equal magnitude, and hence the overall motion may be rectified.} \\

\end{tabular}

 \end{@twocolumnfalse} \vspace{0.6cm}

 ]

\renewcommand*\rmdefault{bch}\normalfont\upshape
\rmfamily
\section*{}
\vspace{-1cm}


\footnotetext{\textit{$^{a}$~Department of Physics, Advanced Materials and Liquid Crystal Institute, Kent State University, Kent, Ohio 44242, USA}}




\section{Introduction}

In nematic liquid crystals, the director field can have many different structures with complex topology.\cite{Mermin1979,Kleman1983,Kleman1989,Kleman2003,Kleman2008,Alexander2012} Some of these structures are defects, such as disclinations, with singularities in the director.\cite{Vromans2016,Tang2017,Long2021} Other structures are nonsingular topological textures, such as skyrmions or knots in the director field, which do not have any singularities but are still unable to relax to a uniform configuration.\cite{Chen2013,Ackerman2014,Leonov2014,Ackerman2017,Afghah2017,Ackerman2017b,Sohn2018,Duzgun2018}

When observed in liquid-crystal cells under a microscope, topological structures often look like particles. Furthermore, these structures appear to move as particles when they are subjected to external forces. For example, Smalyukh and collaborators put skyrmions in a chiral liquid crystal under an applied electric field.\cite{Ackerman2017b,Sohn2018} When the field is switched on, the skyrmions move in a director perpendicular to the induced tilt, and when the field is switched off, the skyrmions move in the opposite direction but with a different magnitude. As a result, if the field is periodically toggled on and off, the skyrmions exhibit a rectified motion, which the experimenters call ``squirming.''

There are already well-established hydrodynamic methods to model the dynamics of liquid crystals. Ericksen-Leslie theory constructs partial differential equations for the director field and the flow velocity field, and then solves these equations to predict the time evolution of the material.\cite{Ericksen1960,Ericksen1961,Leslie1966,Leslie1968} Beris-Edwards theory follows a similar approach, but uses the full tensor field describing the magnitude and director of nematic order, rather than just the director field.\cite{Beris1994} These methods are certainly capable of describing the motion of topological structures.\cite{Toth2002,Svensek2003} Indeed, the squirming motion of skyrmions has already been modeled by such methods.\cite{Ackerman2017b,Sohn2018} Even so, it is still helpful to represent the motion of topological structures on a more coarse-grained basis, like the motion of particles, in order to develop a more intuitive understanding of how they respond to applied forces.

In a recent paper from our group, we developed a coarse-grained theory for the motion of disclinations in two-dimensional (2D) liquid crystals.\cite{Tang2019} In this coarse-grained theory, the basic concept is to represent disclinations by a small number of degrees of freedom, which describe the defect position and orientation. We then express the free energy and the Rayleigh dissipation function in terms of the defect position and orientation. This procedure can be done in two ways: either (a)~construct these functions phenomenologically, based only on symmetry, in terms of the coarse-grained defect position and orientation, or (b)~determine these functions from the Frank free energy and the hydrodynamic dissipation by integrating over the entire director field and flow field. Using both of these approaches, we derive equations of motion for disclinations as effective oriented particles. Shankar \emph{et al.}\ have used a related method to model defect motion in active nematic liquid crystals.\cite{Shankar2018}

The purpose of the current paper is to apply the same concept of coarse-graining to nonsingular topological structures in liquid crystals. In Sec.~2, we begin with the one-dimensional (1D) motion of a sine-Gordon soliton. In this problem, the relevant coarse-grained degrees of freedom are the soliton position and width, and the orientation of the background director field. We determine the free energy and Rayleigh dissipation function in terms of those variables, both by using a phenomenological, symmetry-based approach and by integrating over the entire director field. We thereby obtain equations of motion for the soliton as an effective particle. These equations show explicitly that the soliton moves in response to changing an electric field. If the field is periodically toggled on and off, the motion is rectified because the soliton width changes as a function of field.

In Sec. 3, we extend the argument to the 2D motion of a skyrmion in a chiral liquid crystal. In this case, the important degrees of freedom are the 2D position vector, the radius of the skyrmion, and the three-dimensional (3D) orientation of the background director. We determine the free energy and Rayleigh dissipation function, and then derive the equations of motion. The coarse-grained theory shows that the skyrmion moves in a direction perpendicular to the applied electric field, explicitly because of the chirality. As with the sine-Gordon soliton, the motion is rectified if the field is periodically toggled on and off, because the skyrmion radius changes as a function of field. In those ways, the theory provides simple explanations for key features of the experimental motion.

\section{Sine-Gordon soliton}

\begin{figure}
\centering
\includegraphics[width=\columnwidth]{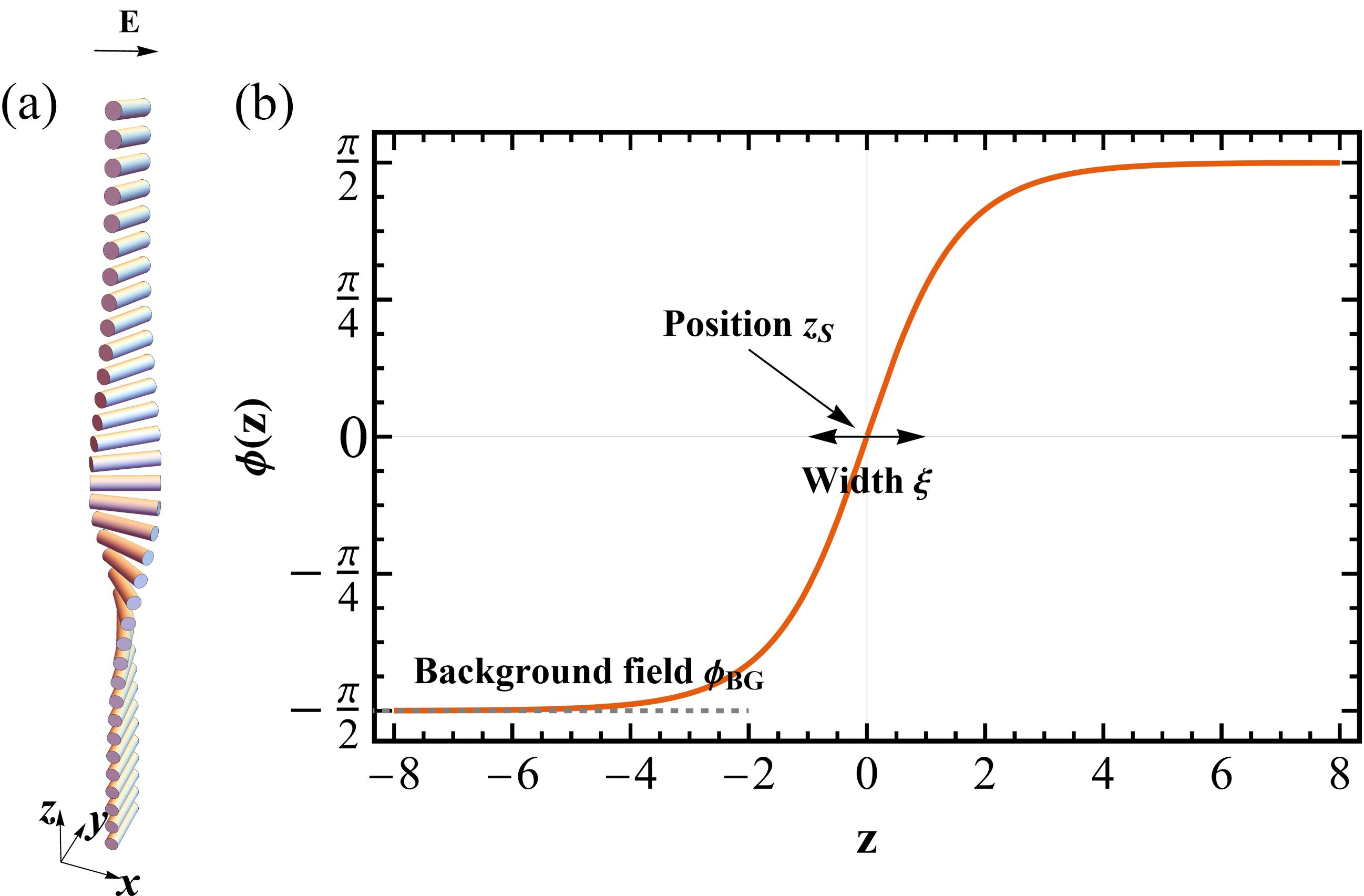}
\caption{(a) Visualization of a soliton in a 1D liquid crystal. The orientation of each cylinder represents the local director. From the bottom to the top, the director twists about the $z$-axis by $\pi$. (b) Plot of the soliton solution for $\phi(z)$ from Eq.~(\ref{solitonsolution}), with the parameters $\phi_E=-\pi/2$, $z_S=0$, and $\xi=1$.}
\label{staticsoliton}
\end{figure}

As a simple example to illustrate the key features of the theoretical approach, we consider the sine-Gordon soliton, shown in Fig.~\ref{staticsoliton}(a). In this example, the director lies in the $(x,y)$ plane, and it varies only as a function of height $z$ and time $t$. Suppose that we apply an electric field, which tends to align the director at a certain background orientation in the $(x,y)$ plane. The director may remain uniform at the favored orientation, or it may exhibit one or more walls, where it twists through an angle of $\pi$. Each of these walls is a sine-Gordon soliton. Each soliton can be regarded as a particle, with a characteristic position and width. The position of the soliton may move up or down in time. However, the soliton cannot just appear or disappear without melting the nematic order, which would cost a prohibitive amount of free energy. In that sense, the soliton is a topological structure.

We first model the soliton using the standard continuum theory, and then consider how to coarse-grain the model.

\subsection{Continuum theory}

In continuum theory, we must represent the director field in the entire liquid crystal as $\hat{\mathbf{n}}(z,t)=(\cos\phi,\sin\phi,0)$, where $\phi(z,t)$ is the azimuthal angle in the $(x,y)$ plane. Because the director $\hat{\mathbf{n}}$ is equivalent to $-\hat{\mathbf{n}}$, the angle $\phi$ is equivalent to $\phi+\pi$. Likewise, we represent the electric field in the $(x,y)$ plane as $\mathbf{E}=(E\cos\phi_E,E\sin\phi_E,0)$. The total free energy per area in the $(x,y)$ plane can then be expressed as
\begin{align}
F &= \int dz \left[\frac{1}{2}K(\partial_i n_j)(\partial_i n_j)
+Kq\hat{\mathbf{n}}\cdot\mathbf{\nabla}\times\hat{\mathbf{n}}
-\frac{1}{8\pi}\epsilon_0\Delta\epsilon(\mathbf{E}\cdot\hat{\mathbf{n}})^2\right]\nonumber\\
&= \int dz \left[\frac{1}{2}K \left(\frac{\partial\phi}{\partial z}\right)^2
-Kq\left(\frac{\partial\phi}{\partial z}\right)
-\frac{\epsilon_0\Delta\epsilon E^2}{8\pi}\cos^2 (\phi-\phi_E)\right].
\label{freeenergytotal}
\end{align}
Here, the first term is the Frank free energy for deformations in the director field, the second term shows the favored twist if the liquid crystal is chiral, and the third term gives the dielectric coupling between the electric field and the director. We consider positive dielectric anisotropy $\Delta\epsilon>0$, so that the director tends to align parallel to the electric field.

To minimize the free energy, we derive the Euler-Lagrange equation in an infinite system
\begin{equation}
K\frac{\partial^2 \phi}{\partial z^2}=\frac{\epsilon_0\Delta\epsilon E^2}{8\pi}\sin2(\phi-\phi_E),
\end{equation}
with the constraint $\phi(+\infty)-\phi(-\infty)=\pi$.
Note that this equation does not involve the chirality $q$, because the chiral term in the free energy is a total derivative. An exact solution is
\begin{equation}
\phi(z)=\phi_E + \frac{\pi}{2} + 2 \arctan\left[\tanh\left(\frac{z-z_S}{\xi}\right)\right],
\label{solitonsolution}
\end{equation}
where
\begin{equation}
\xi=\left(\frac{16\pi K}{\epsilon_0\Delta\epsilon E^2}\right)^{1/2}
\label{solitonwidth}
\end{equation}
is the characteristic width of the soliton, and $z_S$ is any arbitrary position for the soliton center. This solution is plotted in Fig.~\ref{staticsoliton}(b). The free energy of that solution, relative to the free energy of the uniform state $\phi(z)=\phi_E$, is $\Delta F = K(4 \xi^{-1}-\pi q)$. Hence, the soliton has a lower free energy than the uniform state if $q>4/(\pi\xi)$, as discussed previously.\cite{Kamien2001}

The antisoliton, with the opposite sign of twist, has the form $\phi(z)=\phi_E - (\pi/2) - 2 \arctan[\tanh((z-z_S)/\xi)]$. It has a lower free energy than the uniform state if $q<-4/(\pi\xi)$. In an achiral liquid crystal with $q=0$, the soliton and antisoliton each have a higher free energy than the uniform state, but still they are each metastable and cannot relax to the uniform state.

Let us now make the simplest possible model for the dynamics of the liquid crystal, assuming director rotation but no fluid flow. The Rayleigh dissipation function per area in the $(x,y)$ plane can be written as
\begin{equation}
D=\int dz \left[\frac{1}{2}\gamma |\dot{\mathbf{n}}|^2\right]
=\int dz \left[\frac{1}{2}\gamma \dot\phi^2 \right],
\label{energydissipationfunction}
\end{equation}
where $\gamma$ is the rotational viscosity, and $\dot{\mathbf{n}}$ and $\dot\phi$ are the time derivatives. The equation of motion can then be regarded as a balance of forces. The force on $\phi$ from the free energy is
\begin{equation}
- \frac{\delta F}{\delta\phi(z,t)}=K\frac{\partial^2 \phi}{\partial z^2}
-\frac{\epsilon_0\Delta\epsilon E^2}{8\pi}\sin2(\phi-\phi_E),
\end{equation}
and the force on $\phi$ from dissipation is
\begin{equation}
- \frac{\delta D}{\delta\dot\phi(z,t)}=-\gamma\frac{\partial\phi}{\partial t}.
\end{equation}
The sum of these forces must be zero, and hence the equation of motion becomes
\begin{equation}
\gamma\frac{\partial\phi}{\partial t}=K\frac{\partial^2 \phi}{\partial z^2}
-\frac{\epsilon_0\Delta\epsilon E^2}{8\pi}\sin2(\phi-\phi_E).
\label{continuumPDE}
\end{equation}
By solving this partial differential equation (PDE), we can predict the time evolution of the director everywhere in the liquid crystal, both inside and away from the soliton. For example, we can predict how the director responds to a change in the magnitude or direction of the electric field. In principle, this procedure allows us to see how the soliton position moves as a function of time. However, the equation does not show the soliton motion very clearly. For that reason, in the following sections, we consider a more coarse-grained theory.

\subsection{Coarse-grained theory: Phenomenological approach}

At a coarse-grained level, a sine-Gordon soliton is characterized by two variables: the position $\mathbf{r}_S=(0,0,z_s)$ and the width $\xi$. Furthermore, the liquid crystal around the soliton is characterized by the background director $\hat{\mathbf{n}}_{BG}=(\cos\phi_{BG},\sin\phi_{BG},0)$. For a phenomenological theory, we ask: Considering the symmetry of the system, how can the free energy and the Rayleigh dissipation function depend on those three variables?

The free energy must have some term that aligns the background director with the applied electric field. At lowest order in the electric field, this term can be written as $-(\mathbf{E}\cdot\hat{\mathbf{n}}_{BG})^2$, with some arbitrary coefficient. The free energy must also have some term that drives the width $\xi$ toward the favored value $\bar\xi=((16\pi K)/(\epsilon_0\Delta\epsilon E^2))^{1/2}$. This term should diverge for both $\xi\to0$ and $\xi\to\infty$, and hence should be proportional to $\xi+{\bar\xi}^2/\xi$. By contrast, the free energy must be independent of $z_S$, because of translational symmetry along the $z$-axis. Hence, a phenomenological expression for the free energy becomes
\begin{equation}
F=-f_1 E^2 \cos^2 (\phi_{BG}-\phi_E)
+f_2 \left(\xi+\frac{{\bar\xi}^2}{\xi}\right),
\label{freeenergyphenomenological}
\end{equation}
with arbitrary coefficients $f_1$ and $f_2$.

The Rayleigh dissipation function must show the energy dissipation from all of the coarse-grained modes, up to quadratic order in the time derivatives. It must certainly have terms proportional to $|\dot{\mathbf{r}}_S|^2=\dot{z}_S^2$, to $\dot{\xi}^2$, and to $|\dot{\mathbf{n}}_{BG}|^2=\dot{\phi}_{BG}^2$. In addition, it may have a cross term of the form $\dot{\mathbf{r}}_S \cdot(\hat{\mathbf{n}}_{BG}\times\dot{\mathbf{n}}_{BG})=\dot{z}_S\dot{\phi}_{BG}$. Because of the cross product, this term is chiral. It is permitted by symmetry, even if the liquid crystal is not chiral, because the soliton itself is chiral. Combining all of these term, the dissipation function takes the form
\begin{equation}
D=\frac{1}{2}d_1 \dot{\phi}_{BG}^2 + \frac{1}{2}d_2 \dot{z}_S^2 + \frac{1}{2}d_3 \dot{\xi}^2 - d_4\dot{z}_S \dot{\phi}_{BG},
\label{dissipationphenomenological}
\end{equation}
with arbitrary coefficients $d_1$, $d_2$, $d_3$, and $d_4$.

From the free energy and Rayleigh dissipation function, we can derive the equations of motion for the coarse-grained variables. The balance between the force from free energy and the force from dissipation gives
\begin{subequations}
\label{eomforcoarse-grainedvariables}
\begin{align}
-\frac{\partial F}{\partial\phi_{BG}}-\frac{\partial D}{\partial\dot{\phi}_{BG}}&=0,\\
-\frac{\partial F}{\partial z_S}-\frac{\partial D}{\partial\dot{z}_S}&=0,\\
-\frac{\partial F}{\partial\xi}-\frac{\partial D}{\partial\dot{\xi}}&=0.
\end{align}
\end{subequations}
Let us consider Eq.~(\ref{eomforcoarse-grainedvariables}b) for the soliton position. Because the free energy is independent of $z_S$, this equation simplifies to
\begin{equation}
\dot{z}_S = \left(\frac{d_4}{d_2}\right)\dot{\phi}_{BG}.
\label{velocityphenomenological}
\end{equation}
Hence, if the background orientation $\dot{\phi}_{BG}$ rotates, the soliton position $z_S$ must move up or down. In particular, if the electric field rotates in the $(x,y)$ plane with frequency $\omega$, so that $\hat{\mathbf{n}}_{BG}$ also rotates with frequency $\omega$, then the soliton position must move vertically with velocity $(d_4/d_1)\omega$. This motion occurs purely because of the dissipative coupling between soliton velocity and background rotation; it does not require any dependence of free energy on soliton position.

One feature of this phenomenological approach is that it gives very general results, based purely on symmetry, independent of any microscopic details. However, a drawback of this approach is that it does not provide any information about how the arbitrary coefficients depend on more fundamental properties of the liquid crystal. For that information, we must consider an alternative approach.

\subsection{Coarse-grained theory: Integration approach}

For an alternative approach, we want to develop a coarse-grained theory that is explicitly based on the continuum theory of Sec.~2.1. This approach is known as the collective coordinate method in nonlinear dynamics.\cite{McLaughlin1978,Cuevas-Maraver2014} For this calculation, we make an ansatz for time-dependent director field associated with a soliton,
\begin{equation}
\label{solitonansatz}
\phi (z,t) = \phi_{BG}(t) + \frac{\pi}{2}
+ 2 \arctan{\left[ \tanh{\left( \frac{z-z_S(t)}{\xi(t)} \right)} \right]} .
\end{equation}
This ansatz is inspired by the static solution of Eq.~(\ref{solitonsolution}), but now the parameters $\phi_{BG}$, $z_S$, and $\xi$ are allowed to depend on time, and hence the entire director field may depend on time. We put this ansatz into the free energy of Eq.~(\ref{freeenergytotal}) and the Rayleigh dissipation function of Eq.~(\ref{energydissipationfunction}), and express these functions in terms of $\phi_{BG}$, $z_S$, and $\xi$.

In this calculation, we must perform integrals over $z$. Some of these integrals are divergent in the limit of infinite system size. To regularize them, we consider a large box, with $z$ from $-L/2$ to $+L/2$. We assume $L\gg\xi$, and eventually take the limit of $L\to\infty$.

By inserting the ansatz~(\ref{solitonansatz}) into the free energy~(\ref{freeenergytotal}) and integrating over the entire system, we obtain
\begin{equation}
\label{solitonFEansatz}
F=
-\frac{\epsilon_0 \Delta\epsilon E^2 (2L-\xi)}{16\pi} \cos^2 (\phi_{BG}-\phi_E )
+\frac{\epsilon_0 \Delta\epsilon E^2 \xi}{16\pi}
+\frac{2 K}{\xi}.
\end{equation}
In this expression, the first term comes from the dielectric energy of the uniform background director $\hat{\mathbf{n}}_{BG}$ in the electric field $\mathbf{E}$, and it is proportional to the system size $L$. It gives the optimum background director $\phi_{BG}=\phi_E$. The second term is the dielectric energy cost of the soliton, which arises because the director inside the soliton generally does not align with the electric field. This penalty is proportional to the soliton width $\xi$, and hence favors a smaller width. The last term is the elastic energy of the soliton, which is inversely proportional to $\xi$, and hence favors a larger width. The competition between the last two terms gives the optimum width, consistent with Eq.~(\ref{solitonwidth}). Note that this free energy has the form anticipated in Eq.~(\ref{freeenergyphenomenological}) of the phenomenological theory. In particular, it depends on $\phi_{BG}$ and $\xi$, but not on $z_S$, because of translational symmetry along the $z$-axis.

Similarly, by integrating the Rayleigh dissipation function (\ref{energydissipationfunction}) over the entire system, we obtain
\begin{equation}
\label{solitonDFansatz}
D=\frac{L\gamma\dot{\phi}_{BG}^2}{2}+\frac{2\gamma\dot{z}_{S}^2}{\xi}+\frac{\pi^2 \gamma\dot{\xi}^2}{24\xi}-\pi\gamma\dot{z}_{S}\dot{\phi}_{BG}.
\end{equation}
Because the energy dissipation arises from changes of the director field, all of the time derivatives $\dot{\phi}_{BG}$, $\dot{z}_{S}$, and $\dot{\xi}$ contribute to the dissipation function. The first term is associated with changes in the background field, and it is proportional to the system size $L$. The other terms are associated with changes of the soliton position or width, which are localized, and hence those terms are independent of $L$. Interestingly, the last term shows a dissipative coupling between motion of the soliton $\dot{z}_{S}$ and rotation of the background field $\dot{\phi}_{BG}$, as anticipated in Eq.~(\ref{dissipationphenomenological}) of the phenomenological theory. The dissipation function depends on the current value of $\xi$, but not on $\phi_{BG}$ or $z_S$, only on their time derivatives.

Now that we have Eqs.~(\ref{solitonFEansatz}--\ref{solitonDFansatz}) for the integrated free energy and dissipation function, we can put them into Eqs.~(\ref{eomforcoarse-grainedvariables}) for the balance of forces, and derive equations of motion for the three coarse-grained variables. In the limit of large system size $L\gg\xi$, the equation of motion for $\phi_{BG}$ becomes
\begin{equation}
\label{eombackgroundfield}
\gamma\dot{\phi}_{BG}=-\frac{\epsilon_0 \Delta\epsilon E^2}{8\pi}\sin2(\phi_{BG}-\phi_E).
\end{equation}
In this limit of large $L$, the effect of soliton motion on the background field is negligible. Hence, the rotation of $\phi_{\mathrm{BG}}$ is determined only by the external field $\mathbf{E}$, and is independent of the soliton coordinates $z_{S}$, $\dot{z}_{S}$, $\xi$, and $\dot{\xi}$.

In a similar way, the equation of motion for the soliton position $z_{S}$ becomes
\begin{equation}
\label{eomsolitonposition}
\dot{z}_S=\left(\frac{\pi\xi}{4}\right)\dot{\phi}_{BG}.
\end{equation}
This equation of motion is equivalent to Eq.~(\ref{velocityphenomenological}) of the phenomenological theory, but now we see that the arbitrary ratio $d_4/d_2$ has a physical interpretation as $\pi\xi/4$, proportional to the soliton width. Hence, if the electric field rotates with frequency $\omega$, then $\hat{\mathbf{n}}_{BG}$ also rotates with frequency $\omega$, and the soliton position moves vertically with velocity $\pi\xi\omega/4$.

Finally, the equation of motion for the soliton width $\xi$ becomes
\begin{equation}
\label{eomsolitonwidth}
\frac{\pi^2 \gamma}{12\xi}\dot{\xi}=\frac{2K}{\xi^2}
-\frac{\epsilon_0 \Delta\epsilon E^2}{8\pi}\cos2(\phi_{BG}-\phi_E).
\end{equation}
In the next subsection, we will use this equation to explain the rectification of soliton motion in a toggling electric field.

\subsection{Rectified motion}

\begin{figure*}
\centering
\includegraphics[width=\textwidth]{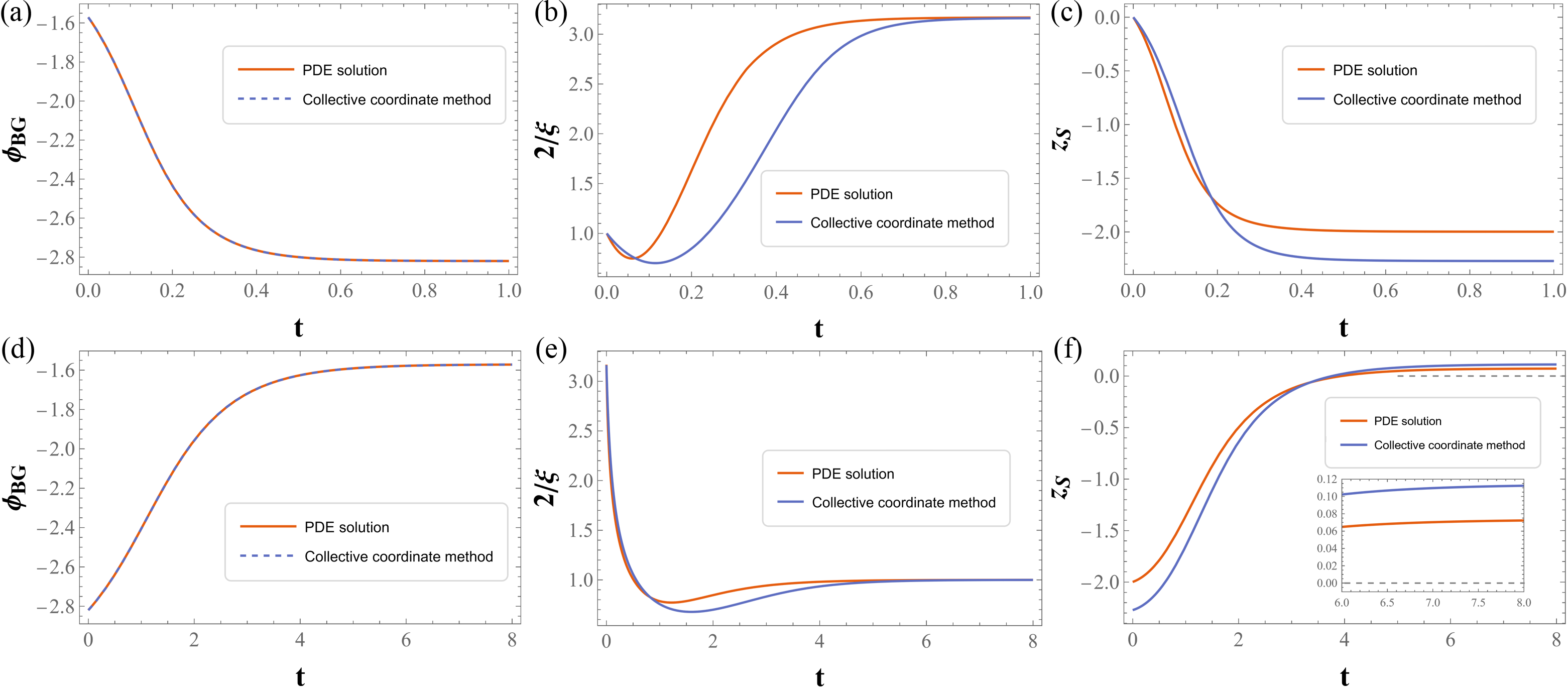}
\caption{Numerical solutions for a 1D soliton in an electric field toggled between $\mathbf{E}^{\mathrm{off}}=(0,1,0)$ and $\mathbf{E}^{\mathrm{on}}=(3,1,0)$. (a)--(c)~Time variation of $\phi_{BG}$, $\xi$ and $z_S$ in the half-period after the electric field is switched on. The soliton starts from a steady state located at $z_S =0$. (d)--(f)~Time variation of these coarse-grained variables in the following half-period after the electric field is switched off. All the numerical solutions of the coarse-grained equations of motion (\ref{eombackgroundfield}), (\ref{eomsolitonposition}), and (\ref{eomsolitonwidth}) are compared with the direct numerical solution of the continuum PDE~(\ref{continuumPDE}).}
\label{nsolutionsforccm}
\end{figure*}

At this point, we would like to assess how well the coarse-grained theory describes the motion of a sine-Gordon soliton. For that purpose, we consider motion induced by a toggling electric field $\mathbf{E}=(E_x(t),E_y,0)$. In this scenario, the static component $E_y$ represents some constant anisotropy, which tends to align the director along the $y$-axis, and the time-varying component $E_x(t)$ represents an applied field that induces director tilt toward the $x$-axis. We choose units such that $\epsilon_0 =4\pi$, $\Delta\epsilon=1$, and $E_y =1$. In these units, we toggle $E_x(t)$ between 0 and 3, with a period of $\tau$, which is chosen to be much greater than the relaxation time of the director field, $\tau\gg\gamma/(\epsilon_0 \Delta\epsilon E^2)$.

We perform numerical calculations in two ways: by solving the continuum PDE~(\ref{continuumPDE}) and by solving the coarse-grained ordinary differential equations (\ref{eombackgroundfield}), (\ref{eomsolitonposition}), and (\ref{eomsolitonwidth}). For the continuum solution, we fit the results to the ansatz of Eq.~(\ref{solitonansatz}), and use the fit to extract the time-dependent parameters $\phi_{BG}(t)$, $\xi(t)$, and $z_S(t)$. Figure~2 shows a comparison of the numerical results from these two solution methods. Parts (a--c) represent the behavior when $E_x$ is switched on, and parts (d--f) represent when the $E_x$ is switched off. The results for $\phi_{BG}(t)$ agree almost perfectly, switching between $\phi_E^\mathrm{off}=\pi/2$ and $\phi_E^\mathrm{on}=\arctan(1/3)\approx0.32$ (mod $\pi$). The results for $\xi(t)$ and $z_S(t)$ generally agree well. The overall agreement shows that the coarse-grained theory is able to capture the basic features of a moving soliton in a changing field. The small discrepancies occur because the continuum solution deviates slightly from the ansatz~(\ref{solitonansatz}) by breaking the odd symmetry about the center of the soliton.

The most important aspect of the numerical results is that the soliton position $z_S(t)$ exhibits a net displacement after the full cycle of field switching. This displacement can be seen in Figs.~2(c) and 2(f): When the field is switched on, the soliton position shifts from $z_S=0$ to a negative value. When the field is switched off again, the soliton position moves back forward and overshoots the initial position, ending with a positive displacement. The net change is small, but it can accumulate over many cycles of the field. Hence, the soliton has a net rectified motion in response to the field toggling. This effect is seen in both the continuum and the coarse-grained solitions.

To explain the rectification, consider the coarse-grained equation of motion~(\ref{eomsolitonposition}) for the soliton position $z_S(t)$. Hypothetically speaking, if the soliton width $\xi$ were independent of time, then we could integrate this equation to obtain $\Delta z_S =(\pi\xi/4)\Delta\phi_{BG}$. For a full cycle of toggling the electric field on and off, the net change in background orientation is $\Delta\phi_{BG}=0$, so the equation would give $\Delta z_S =0$, i.e.\ no rectification. Hence, the observed rectification must be associated with time dependence of the soliton width $\xi$.

The time dependence of $\xi$ can be understood in terms of its dependence on the electric field strength. Equation~(\ref{solitonwidth}) shows that the \emph{equilibrium} soliton width depends inversely on $E$. That equation is not exactly valid during the nonequilibrium dynamics, but still the same trend occurs: A larger field strength leads to a smaller width, and a smaller field strength to a larger width. During the on phase of the cycle, the liquid crystal experiences a stronger field, so $\xi$ is smaller, and the displacement is smaller (in the negative direction because $\Delta\phi_{BG}=\phi_E^\mathrm{on}-\phi_E^\mathrm{off}<0$). During the off phase of the cycle, the liquid crystal experiences a weaker field, so $\xi$ is larger, and the displacement is larger (in the positive direction because $\Delta\phi_{BG}=\phi_E^\mathrm{off}-\phi_E^\mathrm{on}>0$). This difference gives rectification.

To confirm this explanation, we perform a series of calculations for fixed $\phi_E^\mathrm{off}=\pi/2$, $E^\mathrm{off}=1$, $\phi_E^\mathrm{on}=\arctan(1/3)\approx0.32$, and various values of $E^\mathrm{on}$. Figure~3(a) shows numerical results for the net displacement, combining the on and off processes, as a function of $E^\mathrm{on}$. We can see that the net displacement is zero when $E^\mathrm{on}=E^\mathrm{off}$, because the on and off processes are exactly opposite to each other. The net displacement is negative when $E^\mathrm{on}<E^\mathrm{off}$, because the negative displacement in the on process exceeds the positive displacement in the off process. The reverse is true when $E^\mathrm{on}>E^\mathrm{off}$.

Those results for the net displacement correspond to the dynamic trends in the soliton width $\xi$, shown in Fig.~3(b)--3(d). Just before the field is switched on or off, the electric force is balanced by the elastic force, and $\xi$ stays at its equilibrium. Just after the field is switched, the electric force becomes weaker because the equilibrium director has changed to an orientation inside the soliton, while the elastic force remains the same. These unbalanced forces make $\xi$ increase to restore the balance. As the background director gradually shifts to its equilibrium orientation, the electric force becomes stronger, and this change makes $\xi$ decrease until it reaches its equilibrium. For $E^\mathrm{on}<E^\mathrm{off}$ [Fig.~3(b)], the increase and decrease of $\xi$ are exactly opposite in the up and down processes. For $E^\mathrm{on}<E^\mathrm{off}$ [Fig.~3(c)], these trends still occur, but they are not exactly opposite; rather, $\xi$ is generally larger during the on process than during the off process. Conversely, for $E^\mathrm{on}>E^\mathrm{off}$ [Fig.~3(d)], $\xi$ is generally smaller during the on process than during the off process.

\begin{figure}
\centering
\includegraphics[width=\columnwidth]{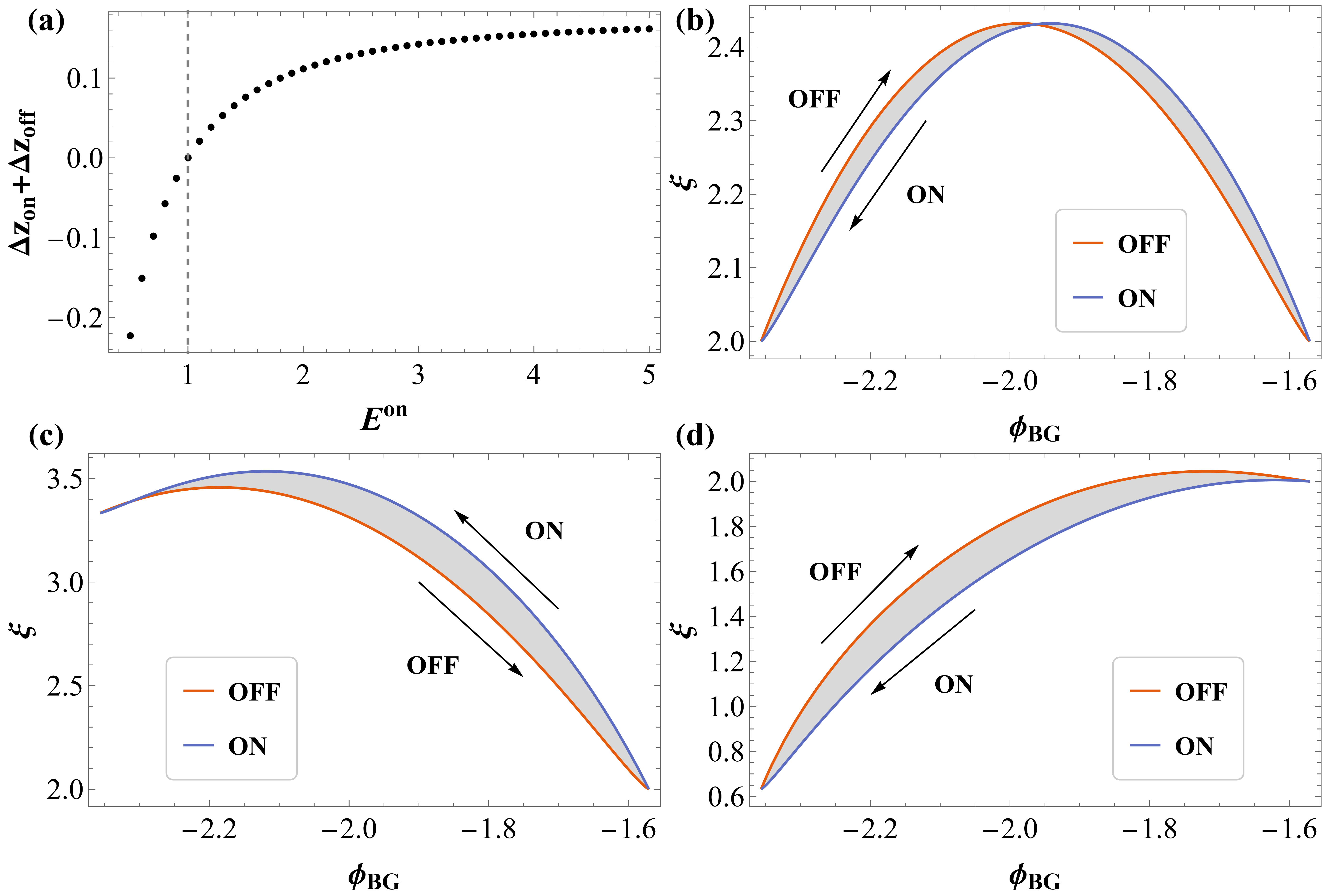}
\caption{Numerical solutions for a 1D soliton in an electric field toggled between $\phi_E^\mathrm{off}=\pi/2$ with $E^\mathrm{off}=1$, and $\phi_E^\mathrm{on}=\arctan(1/3)\approx0.32$ with various values of $E^\mathrm{on}$. (a)~Net displacement of the soliton over one full cycle of toggling, as a function of $E^\mathrm{on}$. (b)--(d)~Soliton width as a function of background orientation $\phi_{BG}$, during the on and off switching processes, with $E^\mathrm{on}=1$, $0.6$, and $10^{1/2}\approx3.16$, respectively. The red lines represent the dynamic change in $\xi$ after the electric field is switched off, and the blue lines represent the corresponding change after the electric field is switched on. The area below each line is proportional to the distance the soliton travels in the corresponding process.}
\label{netdisplacement}
\end{figure}

\section{Skyrmion}

For a further, more complex example of a moving topological structure, we consider a skyrmion, as shown in Fig.~4. In this example, the director varies as a function of $x$ and $y$, and perhaps also time $t$. Far from the center, the director points in the background direction $\hat{\mathbf{n}}_{BG}=\hat{\mathbf{z}}$. Right at the center, the director points in the opposite direction $-\hat{\mathbf{z}}$. In the intermediate region, the director covers all possible orientations on the unit sphere. In a chiral liquid crystal, the director deformation is mainly double twist, and the structure is stabilized because this double twist is compatible with the chirality.

The director field of a skyrmion is well-defined everywhere, with no singularity. Even so, there is no way for the structure to relax to a uniform configuration without melting the nematic order or disrupting the director all the way out to infinity, which would have a prohibitive free energy cost. Hence, like a sine-Gordon soliton, a skyrmion is a nonsingular topological structure.

While a sine-Gordon soliton can move up or down in the $z$ direction, a skyrmion can move in the $(x,y)$ plane. When it moves, the center of the director distortion shifts to another position, while preserving the overall topological properties. Indeed, this motion has been studied experimentally.\cite{Ackerman2017b,Sohn2018} While previous research has modeled the motion through continuum theory, we would like to develop a coarse-grained description, in order to have a simpler understanding of how the skyrmion responds to applied electric fields.

\subsection{Continuum theory}

In principle, continuum theory for a skyrmion is similar to continuum theory for a sine-Gordon soliton. For a skyrmion, we must represent the director field as a three-dimensional unit vector $\hat{\mathbf{n}}(x,y,t)=(\sin\theta\cos\phi,\sin\theta\sin\phi,\cos\theta)$, where $\theta(x,y,t)$ is the polar angle and $\phi(x,y,t)$ is the azimuthal angle. Likewise, the electric field is a 3D vector $\mathbf{E}$. The total free energy per length in the $z$ direction then becomes
\begin{equation}
F = \int dx dy \left[\frac{1}{2}K(\partial_i n_j)(\partial_i n_j)
+Kq\hat{\mathbf{n}}\cdot\mathbf{\nabla}\times\hat{\mathbf{n}}
-\frac{1}{8\pi}\epsilon_0\Delta\epsilon(\mathbf{E}\cdot\hat{\mathbf{n}})^2\right].
\end{equation}
We still assume positive dielectric anisotropy $\Delta\epsilon>0$, so that $\hat{\mathbf{n}}$ tends to align parallel to $\mathbf{E}$.

\begin{figure}
\centering
(a)\includegraphics[width=.44\columnwidth]{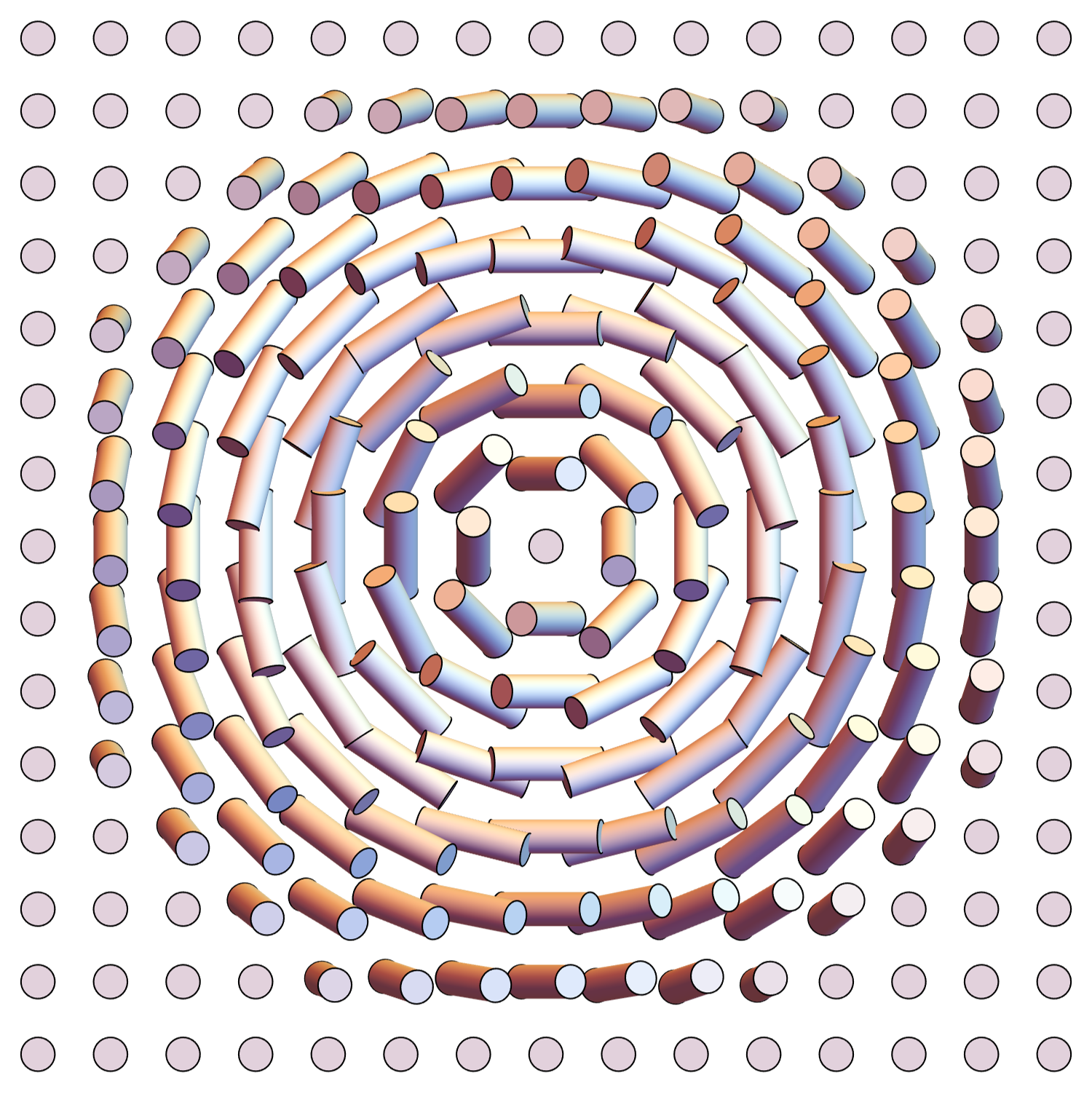}
(b)\includegraphics[width=.44\columnwidth]{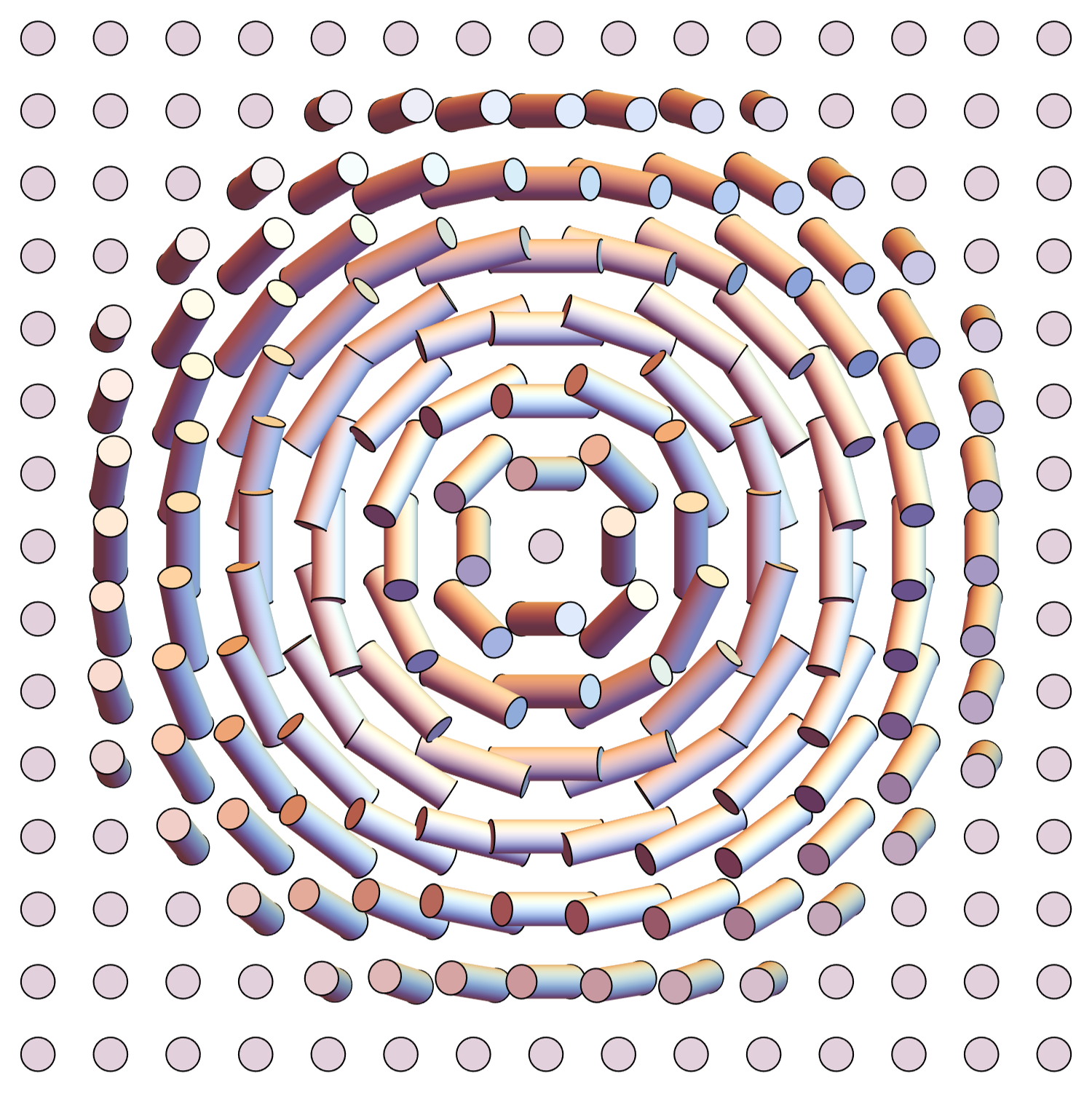}
\caption{Examples of skyrmions in chiral liquid crystals, stabilized by twist. (a)~Right-handed with $\phi_0=\pi/2$. (b) Left-handed with $\phi_0=-\pi/2$.}
\end{figure}

To minimize the free energy, we would like to solve the Euler-Lagrange equations
\begin{equation}
\frac{\delta F}{\delta\theta(x,y,t)}=0, \qquad \frac{\delta F}{\delta\phi(x,y,t)}=0,
\end{equation}
to find a director field with the topology of a skyrmion. Unfortunately, these equations do not have an exact solution. For that reason, researchers have generally used two approaches. First, one can construct an ansatz for the director field, and minimize the free energy over parameters in that ansatz. Alternatively, one can use a purely numerical method to solve the Euler-Lagrange equations for a director field that minimizes the free energy. Either of these approaches can give a skyrmion solution. In the appropriate regime of chirality $q$ and electric field $\mathbf{E}$, the skyrmion has a lower free energy than the uniform state. Outside that regime, even if it has a higher free energy, it is still metastable and cannot relax to the uniform state.

For the dynamics of a skyrmion, we again consider director rotation with no fluid flow. The Rayleigh dissipation function per length in the $z$ direction is
\begin{equation}
D=\int dx dy \left[\frac{1}{2}\gamma |\dot{\mathbf{n}}|^2\right],
\label{dissipationpart3}
\end{equation}
and hence the equations of motion for $\theta$ and $\phi$ become
\begin{equation}
-\frac{\delta F}{\delta\theta(x,y,t)}-\frac{\delta D}{\delta\dot{\theta}(x,y,t)}=0, \quad
-\frac{\delta F}{\delta\phi(x,y,t)}-\frac{\delta D}{\delta\dot{\phi}(x,y,t)}=0.
\end{equation}
These equations can be solved numerically to determine the time evolution of any initial configuration, as the applied electric field is varied.

\subsection{Coarse-grained theory: Phenomenological approach}

Although the continuum theory for a skyrmion is more complex than the continuum theory for a sine-Gordon soliton, we can still use the same phenomenological approach to develop a coarse-grained theory. In this coarse-grained theory, a skyrmion is characterized by its position $\mathbf{r}_S=(x_S,y_S,0)$ and its radius $\xi$. Likewise, the liquid crystal around the skyrmion is characterized by the background director $\hat{\mathbf{n}}_{BG}=(\sin\theta_{BG}\cos\phi_{BG},\sin\theta_{BG}\sin\phi_{BG},\cos\theta_{BG})$. The phenomenological approach considers how the free energy and the Rayleigh dissipation function can depend on those coarse-grained variables.

The free energy must have the form
\begin{equation}
F=-f_1 E^2 \cos^2 (\phi_{BG}-\phi_E)
+f_2 \left(\xi+\frac{\bar{\xi}^2}{\xi}\right),
\end{equation}
analogous to Eq.~(\ref{freeenergyphenomenological}) for the sine-Gordon soliton. In this expression, the first term aligns the background director with the applied electric field. The second term drives the radius $\xi$ toward some arbitrary favored value $\bar\xi$. Because of translational invariance, the free energy must be independent of skyrmion position $\mathbf{r}_S$. The Rayleigh dissipation function must have the form
\begin{equation}
D=\frac{1}{2}d_1 |\dot{\mathbf{n}}_{BG}|^2 + \frac{1}{2}d_2 |\dot{\mathbf{r}}_S|^2 + \frac{1}{2}d_3 \dot{\xi}^2 - d_4 \dot{\mathbf{r}}_S\cdot(\hat{\mathbf{n}}_{BG}\times\dot{\mathbf{n}}_{BG}),
\label{dissipationphenomenologicalskyrmion}
\end{equation}
analogous to Eq.~(\ref{dissipationphenomenological}) for the sine-Gordon soliton. The last term is a chiral dissipative coupling, which is permitted by symmetry because of the chirality of the skyrmion.

Now that we have these expressions for the free energy and Rayleigh dissipation function, we can derive the equations of motion for the coarse-grained variables. In particular, the equations for the skyrmion position are
\begin{equation}
-\frac{\partial F}{\partial x_S}-\frac{\partial D}{\partial\dot{x}_S}=0, \qquad
-\frac{\partial F}{\partial y_S}-\frac{\partial D}{\partial\dot{y}_S}=0.
\end{equation}
Because the free energy is independent of $x_S$ and $y_S$, these equations simplify to
\begin{equation}
\dot{x}_S = \left(\frac{d_4}{d_2}\right)(\hat{\mathbf{n}}_{BG}\times\dot{\mathbf{n}}_{BG})_x , \qquad
\dot{y}_S = \left(\frac{d_4}{d_2}\right)(\hat{\mathbf{n}}_{BG}\times\dot{\mathbf{n}}_{BG})_y .
\end{equation}
From these equations, we can see that the skyrmion moves in the direction given by $\hat{\mathbf{n}}_{BG}\times\dot{\mathbf{n}}_{BG}$, projected into the $(x,y)$ plane. For example, if $\hat{\mathbf{n}}_{BG}$ is initially in the $z$ direction, and it rotates toward the $x$ direction, then the skyrmion moves in the $y$ direction. This motion is a direct result of the chirality of the skyrmion, analogous to the motion of a sine-Gordon soliton discussed earlier. To estimate the magnitude of the velocity, we need to use the integration approach to connect the phenomenological parameters with smaller-scale features of the skyrmion structure.

\subsection{Coarse-grained theory: Integration approach}

To implement the integration approach, we must construct an ansatz for the director field of a skyrmion, insert that ansatz into the continuum free energy and Rayleigh dissipation function, and integrate over the entire system. Here, we are mainly interested in the dynamic behavior of the skyrmion position $\mathbf{r}_S$, which should depend only on the dissipation function, not the free energy, because of translation invariance. Hence, we concentrate on the dissipation function rather than the free energy.

The ansatz for $\hat{\mathbf{n}}(\mathbf{r})$ must depend on the skyrmion position $\mathbf{r}_S$ and radius $\xi$, as well as the background director $\hat{\mathbf{n}}_{BG}$, and hence we write it as $\hat{\mathbf{n}}(\mathbf{r};\mathbf{r}_S,\xi,\hat{\mathbf{n}}_{BG})$. Previous theoretical research has normally studied the case of a skyrmion centered at the origin, $\mathbf{r}_S =0$, with a background director in the vertical direction, $\hat{\mathbf{n}}_{BG}=\hat{\mathbf{z}}$. In that case, the director field is most conveniently written in cylindrical coordinates $(\rho,\phi)$ as the hedgehog ansatz
\begin{equation}
\hat{\mathbf{n}}(\mathbf{r};0,\xi,\hat{\mathbf{z}}) =
\begin{pmatrix}
 \sin{f(\rho/\xi)}\cos{(\phi+\phi_0)} \\
 \sin{f(\rho/\xi)}\sin{(\phi+\phi_0)} \\
 \cos{f(\rho/\xi)}
\end{pmatrix}.
\label{hedgehog}
\end{equation}
Here, the function $f(\rho/\xi)$ is the position-dependent polar angle, with the limits $f(0)=\pi$ and $f(\infty)=0$; it generally decays exponentially for large $\rho$. The constant $\phi_0 =\pm\pi/2$ represents the right- or left-handed structure of the skyrmion; one handedness is favored by the free energy in a chiral liquid crystal, and the other is disfavored.

If the skyrmion is not centered at the origin, then we can easily modify this ansatz with a translation by the vector $\mathbf{r}_S$, giving
\begin{equation}
\hat{\mathbf{n}}(\mathbf{r};\mathbf{r}_S,\xi,\hat{\mathbf{z}}) =
\hat{\mathbf{n}}(\mathbf{r}-\mathbf{r}_S;0,\xi,\hat{\mathbf{z}}).
\end{equation}
If the background director $\hat{\mathbf{n}}_{BG}$ is not vertical, then it can be related to the vertical direction through a rotation $\hat{\mathbf{n}}_{BG}=\mathbf{R}\cdot\hat{\mathbf{z}}$, where $\mathbf{R}$ is the matrix representing a rotation through the angle $\theta_{BG}$ about the axis $(-\sin\phi_{BG},\cos\phi_{BG},0)$. Hence, we apply the same rotation matrix to the entire director field, which gives the full ansatz
\begin{equation}
\label{rotationtransform}
\hat{\mathbf{n}}(\mathbf{r};\mathbf{r}_S,\xi,\hat{\mathbf{n}}_{BG}) = \mathbf{R}\cdot\hat{\mathbf{n}}(\mathbf{r};\mathbf{r}_S,\xi,\hat{\mathbf{z}}).
\end{equation}
In constructing this rotation, we implicitly assume that $\hat{\mathbf{n}}_{BG}$ is close to $\hat{\mathbf{z}}$, and $-\hat{\mathbf{n}}_{BG}$ is close to $-\hat{\mathbf{z}}$, not vice versa. Hence, the results will be symmetric under simultaneously changing the signs of $\hat{\mathbf{n}}_{BG}$ and $\hat{\mathbf{z}}$, but not under separately changing the signs of $\hat{\mathbf{n}}_{BG}$ or $\hat{\mathbf{z}}$.

We now insert the ansatz into the dissipation function of Eq.~(\ref{dissipationpart3}) to obtain
\begin{align}
\label{skyrmiondissipation}
D&=\frac{1}{2}\gamma\int dx dy \left|
\dot{r}_{Si}\frac{\partial\hat{\mathbf{n}}}{\partial r_{Si}}
+\dot{\xi}\frac{\partial\hat{\mathbf{n}}}{\partial\xi}
+\dot{n}_{BGi}\frac{\partial\hat{\mathbf{n}}}{\partial n_{BGi}}
\right|^2\\
&=\frac{1}{2}\gamma\dot{r}_{Si}\dot{r}_{Sj}\int dx dy \frac{\partial n_k}{\partial r_{Si}}\frac{\partial n_k}{\partial r_{Sj}}
+\gamma\dot{r}_{Si}\dot{\xi}\int dx dy \frac{\partial n_k}{\partial r_{Si}}\frac{\partial n_k}{\partial\xi}\nonumber\\
&\qquad+\gamma\dot{r}_{Si}\dot{n}_{BGj}\int dx dy \frac{\partial n_k}{\partial r_{Si}}\frac{\partial n_k}{\partial n_{BGj}}
+\text{terms independent of } \dot{\mathbf{r}}_S .\nonumber
\end{align}
The first term simplifies to $D^{(1)}=\frac{1}{2}C\gamma |\dot{\mathbf{r}}_S |^2$, where
\begin{equation}
C=\pi\int_0^\infty u du \left(\left(\frac{\partial f(u)}{\partial u}\right)^2+\frac{\sin^2 f(u)}{u^2}\right)
\label{integralC}
\end{equation}
is a dimensionless constant of order 1, which depends on the form of $f(\rho/\xi)$. This term shows the isotropic drag on a moving skyrmion. The second term, involving $\dot{\mathbf{r}}_S$ and $\dot{\xi}$, integrates to $D^{(2)}=0$, which indicates that a change of skyrmion radius does not drive a skyrmion to move.

The third term shows a dissipative coupling between $\dot{\mathbf{r}}_S$ and $\dot{\mathbf{n}}_{BG}$. After some long calculations, it simplifies to
\begin{align}
D^{(3)}&=\pm C'\gamma\xi\dot{\mathbf{r}}_S \cdot \biggl[(\hat{\mathbf{n}}_{BG}\times\dot{\mathbf{n}}_{BG})
+(1-\hat{\mathbf{z}}\cdot\hat{\mathbf{n}}_{BG})(\hat{\mathbf{z}}\times\dot{\mathbf{n}}_{BG})\nonumber\\
&\qquad\qquad\qquad+\frac{(\hat{\mathbf{z}}\cdot\hat{\mathbf{n}}_{BG})(\hat{\mathbf{z}}\cdot\dot{\mathbf{n}}_{BG})
(\hat{\mathbf{z}}\times\hat{\mathbf{n}}_{BG})}{1+\hat{\mathbf{z}}\cdot\hat{\mathbf{n}}_{BG}}\biggr],
\label{dissipationchirallong}
\end{align}
where
\begin{equation}
C'=\pi\int_0^\infty du\left(u\frac{\partial f(u)}{\partial u}+\frac{\sin(2f(u))}{2}\right)
\label{integralCprime}
\end{equation}
is another dimensionless constant of order 1. In Eq.~(\ref{dissipationchirallong}), the $\pm$ sign corresponds to the chirality $\phi_0=\pm\pi/2$ in the ansatz. For $\hat{\mathbf{n}}_{BG}$ close to $\hat{\mathbf{z}}$, or $\theta_{BG}\ll1$, the first term in square brackets scales as $\theta_{BG}^1$, while the second and third terms scale as $\theta_{BG}^3$. Hence, we neglect the higher-order terms, and consider only
$D^{(3)} = {\pm C'\gamma\xi\dot{\mathbf{r}}_S \cdot (\hat{\mathbf{n}}_{BG}\times\dot{\mathbf{n}}_{BG})}$.

Combining these components, the total dissipation function becomes
\begin{equation}
D=\frac{1}{2}C\gamma |\dot{\mathbf{r}}_S |^2
\pm C'\gamma\xi\dot{\mathbf{r}}_S \cdot (\hat{\mathbf{n}}_{BG}\times\dot{\mathbf{n}}_{BG})
+\text{terms independent of } \dot{\mathbf{r}}_S .
\end{equation}
Hence, the equations of motion for the soliton position become
\begin{equation}
\dot{x}_S = \mp\left(\frac{C'\xi}{C}\right)(\hat{\mathbf{n}}_{BG}\times\dot{\mathbf{n}}_{BG})_x , \qquad
\dot{y}_S = \mp\left(\frac{C'\xi}{C}\right)(\hat{\mathbf{n}}_{BG}\times\dot{\mathbf{n}}_{BG})_y .
\end{equation}
These expressions have the same form that was anticipated in the phenomenological theory, based only on symmetry considerations, but now we have explicit expressions for the coefficients in terms of the soliton width $\xi$ and dimensionless factors $C$ and $C'$.

\subsection{Rectified motion}

The results of this section for skyrmions moving in 2D are quite analogous to the results of the previous section for sine-Gordon solitons moving in 1D. In each case, the motion of the topological structure is controlled by a dissipative coupling with the background director, and the background director is controlled by the applied electrici field. We can use this analogy to predict the results of a field-toggling experiment on a skyrmion.

Suppose that we apply an electric field that causes $\hat{\mathbf{n}}_{BG}$ to tilt from $\hat{\mathbf{z}}$ toward $\hat{\mathbf{x}}$. In that case, $\dot{\mathbf{n}}_{BG}$ is in the $x$ direction, so $\hat{\mathbf{n}}_{BG}\times\dot{\mathbf{n}}_{BG}$ is in the $y$ direction. Hence, this tilt leads the skyrmion to move in the $y$ direction. Likewise, tilting the background director toward the $y$ direction leads the skyrmion to move in the $-x$ direction. In either case, the velocity of skyrmion motion scales as the rotational velocity of $\hat{\mathbf{n}}_{BG}$ times the skyrmion radius $\xi$. Bringing $\hat{\mathbf{n}}_{BG}$ back to $\hat{\mathbf{z}}$ leads to skyrmion motion in the opposite direction. As with the sine-Gordon soliton, the magnitude of reverse motion should differ from the magnitude of forward motion, if the radius $\xi$ changes due to a change in the applied electric field magnitude. Hence, a single cycle of the applied field should give some net motion, and repeated cycles should give rectified motion perpendicular to the tilt direction.

The scenario suggested here is similar to the experiments of Smalyukh and collaborators, which study skyrmion motion under a toggling electric field.\cite{Ackerman2017b,Sohn2018} In those experiments, the role of $E_z$ is played by a static anisotropy in a thin cell, resulting from surface anchoring, which tends to align $\hat{\mathbf{n}}_{BG}$ in the vertical direction. The role of $E_x$ is played by a vertical electric field acting on a liquid crystal with \emph{negative} dielectric anisotropy, which tends to push $\hat{\mathbf{n}}_{BG}$ toward the $(x,y)$ plane, together with a surface pretilt that favors the $x$ direction within that plane. These experiments show that tilting $\hat{\mathbf{n}}_{BG}$ from the $z$ direction toward the $x$ direction leads to skyrmion motion in the $y$ direction. Likewise, bringing $\hat{\mathbf{n}}_{BG}$ back toward $z$ leads to skyrmion motion in the opposite direction, but with a different magnitude. Repeated cycles of the field thus give net rectified motion, which the experimenters describe as ``squirming.''

Of course, we recognize that the experiments are more complex than the theoretical scenario. In particular, the experiments involve a 3D liquid cell with anchoring conditions on the top and bottom surfaces, rather than the idealized 2D liquid crystal considered here. The experimental group has already done detailed theoretical modeling of the 3D cell under field toggling, and the modeling agrees with the observed behavior. Our current work certainly does not match the level of detail in that modeling. We only suggest that our work identifies the minimal theoretical features that are needed to explain the observed squirming motion: the dissipative coupling between skyrmion motion and the background director, arising from the chirality of the skyrmions, and the change in the skyrmion size between the forward and reverse processes. In that way, it identifies the types of topological structures that should exhibit similar dynamic behavior.

\section{Conclusions}

The research in this paper leads to both specific conclusions for the dynamics of skyrmions and general conclusions for theoretical methods.

For skyrmions, our main conclusion is that their motion can be described as the dynamics of effective particles, in spite of the fact that they are really topological structures of the director field. Skyrmion motion has conventional isotropic drag, and it also has an interesting dissipative coupling with rotation of the background director field. The particle-like description of skyrmion dynamics is similar to the particle-like description of disclination dynamics in previous studies.\cite{Tang2019,Shankar2018} Indeed, the particle-like description works even better for skyrmions than for disclinations, because the integrals~(\ref{integralC}) and (\ref{integralCprime}) for effective drag coefficients are convergent in the skyrmion case, while the analogous integrals diverge with the system size in the disclination case.

More generally, we have used a theoretical approach based on the free energy and the Rayleigh dissipation function, and this theoretical approach seems to be promising for describing dissipative dynamics of \emph{any} localized structure. One helpful feature of this approach is that it can be done on any length scale: We can construct the free energy and Rayleigh dissipation function in terms of mesoscopic order parameter fields, such as the nematic director field, or in terms of more macroscopic variables, such as the positions or orientations of topological structures. By taking appropriate derivatives of those functions, we can find the forces and derive the equations of motion for mesoscopic order parameters fields or macroscopic variables. If we wish to develop a coarse-grained theory in terms of macroscopic variables, there are two options: We can construct a phenomenological theory based purely on symmetry considerations, which gives quick results in terms of arbitrary coefficients, or we can use the integration method to derive the coefficients in terms of known parameters in the continuum theory. This coarse-grained theoretical approach should be useful for further studies of the dynamics of many different topological structures.

\section*{Conflicts of interest}

There are no conflicts to declare.

\section*{Acknowledgements}

This work was supported by National Science Foundation Grant No.~DMR-1409658.

\section*{Appendix: Skyrmions stabilized by splay (Rashba) rather than twist (Dresselhaus)}

\begin{figure}
\centering
(a)\includegraphics[width=.44\columnwidth]{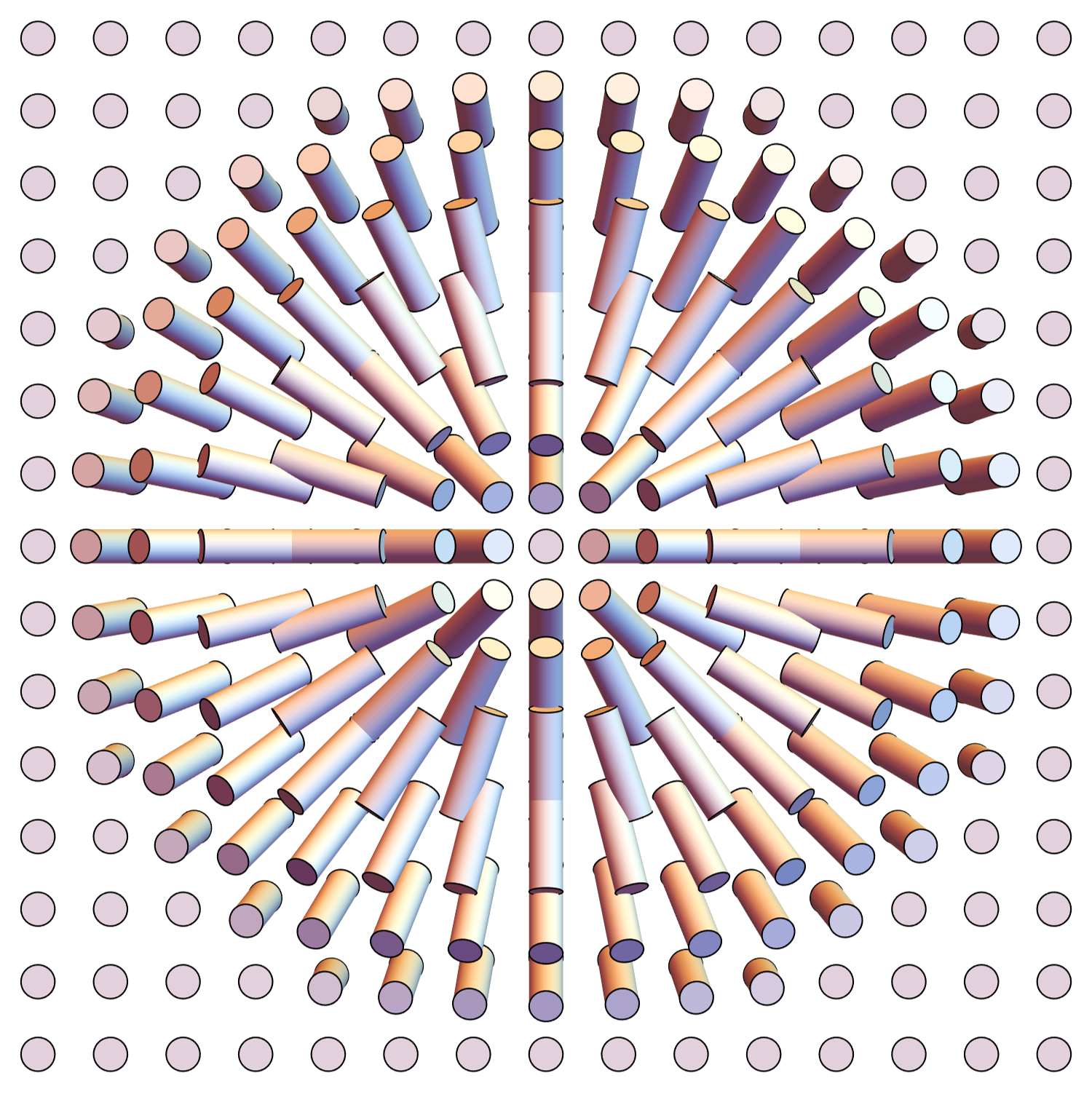}
(b)\includegraphics[width=.44\columnwidth]{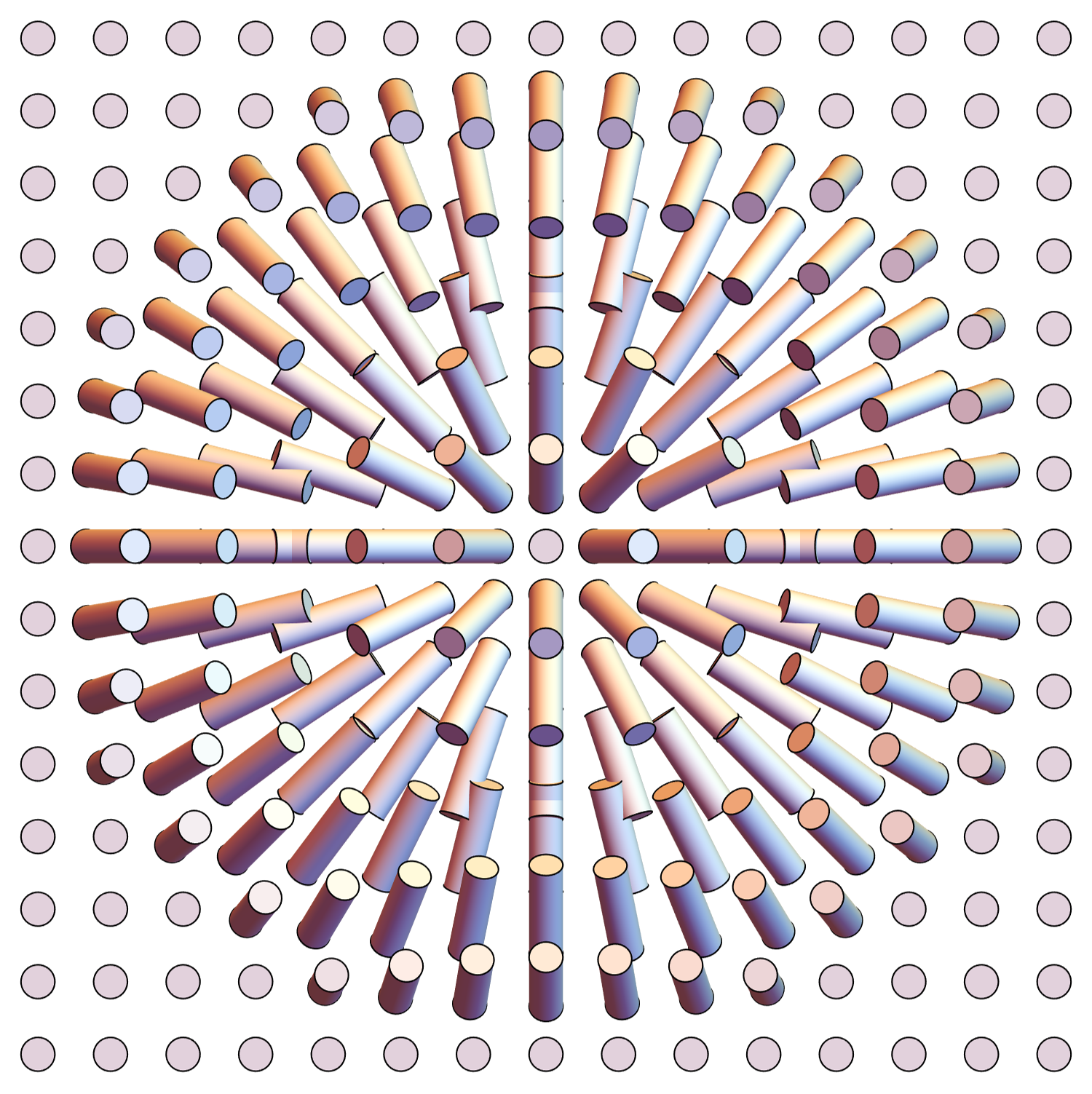}
\caption{Examples of skyrmions stabilized by splay. (a)~Parameter $\phi_0=0$. (b) Parameter $\phi_0=\pi$.}
\end{figure}

In this article, we have discussed one mechanism to stabilize skyrmions, which is the favored twist arising from chirality. However, that is not the only possible mechanism. Another possibility is the favored splay arising from polarity---either a spontaneous polarity in the bulk or an induced polarity at the surface of a liquid crystal. Long ago, Meyer and Pershan found experimentally and theoretically that surface polarity can induce striped domains in liquid crystals.\cite{Meyer1973} A more recent perspective on mechanisms for director modulations has been presented by our group.\cite{Selinger2021}

In the liquid crystal literature, to our knowledge, experiments have only reported twist-stabilized skyrmions, not splay-stabilized skyrmions. However, in the literature on magnetic skyrmions, both types of skyrmions have been discussed.\cite{Rowland2016} In that literature, the twist mechanism is called ``Dresselhaus spin-orbit coupling,'' and the splay mechanism is called ``Rashba spin-orbit coupling.'' It is possible that splay-stabilized skyrmions will also be found in liquid crystals. Such skyrmions would have the structure shown in Fig.~5.

In this appendix, we show how the results in this article about coarse-grained dynamics are modified for splay-stabilized rather than twist-stabilized skyrmions.

First, consider the phenomenological approach. For splay-stabilized skyrmions in an achiral liquid crystal, symmetry does not permit the cross product term $\dot{\mathbf{r}}_S\cdot(\hat{\mathbf{n}}_{BG}\times\dot{\mathbf{n}}_{BG})$ in the Rayleigh dissipation function of Eq.~(\ref{dissipationphenomenologicalskyrmion}). However, the liquid crystal does have polarity $\mathbf{P}$ along to the director, so its dissipation function can have a term proportional to $(\dot{\mathbf{r}}_S \cdot\dot{\mathbf{n}}_{BG})(\mathbf{P}\cdot\hat{\mathbf{n}}_{BG})$. If $\hat{\mathbf{n}}_{BG}$ and $\mathbf{P}$ are both close to $\hat{\mathbf{z}}$ direction, this term simplifies to $(\dot{\mathbf{r}}_S \cdot\dot{\mathbf{n}}_{BG})$, so the full dissipation function becomes
\begin{equation}
D=\frac{1}{2}d_1 |\dot{\mathbf{n}}_{BG}|^2 + \frac{1}{2}d_2 |\dot{\mathbf{r}}_S|^2 + \frac{1}{2}d_3 \dot{\xi}^2 - d'_4 (\dot{\mathbf{r}}_S \cdot\dot{\mathbf{n}}_{BG}).
\label{dissipationsplayskyrmion}
\end{equation}
Because the free energy is still independent of the skyrmion position, the equations of motion become simply $\partial D/\partial\dot{x}_S = \partial D/\partial\dot{y}_S = 0$, so $\dot{x}_S=(d'_4/d_1)(\dot{\mathbf{n}}_{BG})_x$ and $\dot{y}_S=(d'_4/d_1)(\dot{\mathbf{n}}_{BG})_y$. Hence, the splay-stabilized skyrmion moves parallel to $\dot{\mathbf{n}}_{BG}$, projected into the $(x,y)$ plane, unlike the twist-stabilized skyrmion, which moves perpendicular to $\dot{\mathbf{n}}_{BG}$, projected into the $(x,y)$ plane.

Next, consider the integration approach. In the hedgehog ansatz of Eq.~(\ref{hedgehog}), twist-stabilized skyrmions have the parameter $\phi_0=\pm\pi/2$, corresponding to the right- or left-handed structures in Fig.~4. By contrast, splay-stabilized skyrmions have the parameter $\phi_0 = 0$ or $\pi$, corresponding to the inward or outward splay structures in Fig.~5. Hence, we repeat the integration of the dissipation function for general $\phi_0$, which gives
\begin{align}
D&=\frac{1}{2}C\gamma |\dot{\mathbf{r}}_S |^2
+C'\gamma\xi\dot{\mathbf{r}}_S \cdot
\Biggl\{\biggl[(\hat{\mathbf{n}}_{BG}\times\dot{\mathbf{n}}_{BG})
+(1-\hat{\mathbf{z}}\cdot\hat{\mathbf{n}}_{BG})(\hat{\mathbf{z}}\times\dot{\mathbf{n}}_{BG})\nonumber\\
&\qquad+\frac{(\hat{\mathbf{z}}\cdot\hat{\mathbf{n}}_{BG})(\hat{\mathbf{z}}\cdot\dot{\mathbf{n}}_{BG})
(\hat{\mathbf{z}}\times\hat{\mathbf{n}}_{BG})}{1+\hat{\mathbf{z}}\cdot\hat{\mathbf{n}}_{BG}}\biggr]
\sin\phi_0\\
&\qquad-\biggl[\dot{\mathbf{n}}_{BG}
-\frac{\hat{\mathbf{z}}\cdot\dot{\mathbf{n}}_{BG}}{1+\hat{\mathbf{z}}\cdot\hat{\mathbf{n}}_{BG}}\biggr]\cos\phi_0\Biggr\} +\text{terms independent of } \dot{\mathbf{r}}_S .\nonumber
\end{align}
For splay-stabilized skyrmions, that expression reduces to
\begin{align}
D&=\frac{1}{2}C\gamma |\dot{\mathbf{r}}_S |^2
\mp C'\gamma\xi\dot{\mathbf{r}}_S \cdot \biggl[\dot{\mathbf{n}}_{BG}
-\frac{\hat{\mathbf{z}}\cdot\dot{\mathbf{n}}_{BG}}{1+\hat{\mathbf{z}}\cdot\hat{\mathbf{n}}_{BG}}\biggr]\nonumber\\
&\qquad+\text{terms independent of } \dot{\mathbf{r}}_S ,
\end{align}
with the upper or lower sign corresponding to $\phi_0 = 0$ or $\pi$. For $\hat{\mathbf{n}}_{BG}$ close to $\hat{\mathbf{z}}$, or $\theta_{BG}\ll1$, the first term in square brackets scales as $\theta_{BG}^1$, while the second term scales as $\theta_{BG}^3$. Hence, we neglect the higher-order term, and obtain a dissipation function consistent with Eq.~(\ref{dissipationsplayskyrmion}) anticipated from the phenomenological approach.

Based on either the phenomenological or the integration approach, we see the dissipation function is quite similar for splay-stabilized skyrmions and for twist-stabilized skyrmions. However, the key difference is that the velocity $\dot{\mathbf{r}}_S$ is coupled with $\dot{\mathbf{n}}_{BG}$ for splay-stabilized skyrmions, while it is coupled with $\hat{\mathbf{n}}_{BG}\times\dot{\mathbf{n}}_{BG}$ for twist-stabilized skyrmions. This difference should have a direct consequence for motion in a field-toggling experiment: Splay-stabilized skyrmions should exhibit a ``squirming'' rectified motion in the direction aligned with the tilt change $\dot{\mathbf{n}}_{BG}$, while twist-stabilized skyrmions show this motion perpendicular to the tilt change. This result shows the power of coarse-grained theory to predict motion based on symmetry considerations, without detailed simulations.



\balance


\bibliography{solitonmotion3} 
\bibliographystyle{rsc} 

\end{document}